\documentclass[prd,showpacs]{revtex4}
\usepackage{amssymb}
\usepackage{amsmath}
\usepackage{color}
\usepackage[colorlinks=true,linkcolor=red]{hyperref}
\usepackage{hyperref}
\begin{document}

\title{New $3$-Brane Solutions in $5D$ Spacetime}

\author{Pavle Midodashvili${}^{1}$}
\email{pmidodashvili@yahoo.com}
\author{Levan Midodashvili${}^{2}$}
\email{levmid@hotmail.com}

\affiliation{${\ }^{(1)}
$Ilia State University, Kakutsa Cholokashvili Ave 3/5, Tbilisi 0162, Georgia \\
${\ }^{(2)}$Gori University, Chavchavadze str. 53, Gori 1400, Georgia}

%\date{\today}
\date{19 October 2010}

\begin{abstract}
In the article we consider the 5D spacetime physical set-up with a ghost-like bulk scalar field and the 3-brane at the origin of extra coordinate. Performing detailed investigation of corresponding action and Einstein equations we present new possible brane model solutions.
\end{abstract}


\maketitle

%%%%%%%%%%%%%%%%%%%%%%%%%%%%%%%%%%%%%%%%%%%%%%%%%%%%%%%%%%%%%%%

Recently in the paper \cite{gog-sing_2010} it was presented a simple standing wave brane model in 5D spacetime.
Corresponding physical system consists of 3-brane located at the origin of extra dimension and a phantom/ghost scalar field in the bulk.

In  this article for the same physical set-up we present new possible solutions that dramatically differ from those found in the paper \cite{gog-sing_2010}.

In what follows we will derive new brane model solutions by consistent and detailed investigation of corresponding action and Einstein equations of the system.

Let's begin with action. The 5D action of the system under consideration is
\begin{equation}\label{Action}
S = \frac{1}{{2{k^2}}}\int {{d^5}x\sqrt g \left\{ {\left[ {R - 2\Lambda} \right] + {L_{{\rm{Ghost}}}} + {L_{{\rm{Br}}}}} \right\}}~,
\end{equation}
 where parameter $k$ obeys the relation ${k^2} = 8\pi G = \frac{{8\pi }}{{{M^3}}}$  ($G$ and $M$ are the 5-dimensional Newton constant and the 5-dimensional Planck mass scale, respectively), ${L_{{\rm{Br}}}}$ is the 3-brane Lagrangian, $L_{{\rm{Ghost}}}$ is the phantom/ghost field Lagrangian and $\Lambda$ is 5D cosmological constant.

The variation of the action \eqref{Action} with respect to the 5-dimensional metric tensor ${g_{AB}}$ leads to Einstein equations:

\begin{equation}\label{EinsteinEquations1}
{R_{AB}} - \frac{1}{2}{g_{AB}}R =  -  \Lambda{g_{AB}} + {k^2}{T_{AB}}+{k^2}{\lambda_{AB}}.
\end{equation}
 Here $T_{AB}$ is phantom/ghost scalar field energy-momentum tensor
 \begin{equation}\label{GhostEnergyMomentumTensor}
{T_{AB}} =  - {\partial _A}\phi {\partial _B}\phi  + \frac{1}{2}{g_{AB}}{\partial ^C}\phi {\partial _C}\phi,
\end{equation}
and $\lambda_{AB}$ denotes the energy-momentum tensor of the brane which acts as a gravitational source even in the absence of particle excitations on it.

Using \eqref{GhostEnergyMomentumTensor} equations \eqref{EinsteinEquations1} can be rewritten in the form:
\begin{equation}\label{EinsteinEquations2}
R_{AB}=\frac{2}{3}\Lambda g_{AB} - k^2 \partial_A \phi \partial_B \phi+k^2 \left(\lambda_{AB}-\frac{1}{3}g_{AB}\lambda\right)~,
\end{equation}
where $\lambda=g^{AB}\lambda_{AB}$.

The metric ansatz we use is
\begin{equation}\label{MetricAnsatz}
d{s^2} = {e^{2a|r|}}\left[ {d{t^2} - {e^{u\left( {t,|r|} \right)}}\left( {d{x^2} + d{y^2}} \right) - {e^{ - 2u\left( {t,|r|} \right)}}d{z^2}} \right] - d{r^2}~,
\end{equation}
where $a$ is a constant and the function $u$ depends on time $t$ and on absolute value $|r|$ of extra coordinate which varies within $\left]-\infty,+\infty\right[$. In the limiting case $\phi \equiv 0$ and $u(t,|r|) \equiv 0$ the model will be identical to the well known Randall-Sundrum model \cite{Randall-Sundrum}.

Taking into account symmetry properties of the metric \eqref{MetricAnsatz} we assume  that phantom-like scalar field depends  only on time $t$ and absolute value $|r|$ of extra coordinate, i.e. $\phi=\phi\left(t,|r|\right)$.

Due to all conditions introduced above equations \eqref{EinsteinEquations2} will have the following explicit form:

\begin{equation}
\begin{array}{l}
 - \frac{3}{2}{\left( {\frac{{\partial u}}{{\partial t}}} \right)^2} + 4{a^2}{e^{2a\left| r \right|}} + 2a\delta \left( r \right){e^{2a\left| r \right|}} = \frac{2}{3}\Lambda {e^{2a\left| r \right|}} - {k^2}{\left( {\frac{{\partial \phi }}{{\partial t}}} \right)^2} + \\ +\frac{1}{3}{k^2}\left( {{2\lambda _{tt}} + {e^{- u}}{\lambda _{xx}} + {e^{- u}}{\lambda _{yy}} + {e^{2u}}{\lambda _{zz}} + {e^{ 2a|r|}}{\lambda _{rr}}} \right), \\
\nonumber \\
\frac{1}{2}{e^u}\left( {\frac{{{\partial ^2}u}}{{\partial {t^2}}} - {e^{2a|r|}}\left( {8{a^2} + 4a \frac{{\partial u}}{{\partial {|r|}}}+2\delta \left( r \right)\frac{{\partial u}}{{\partial {|r|}}}  + 4a\delta \left( r \right) + \frac{{{\partial ^2}u}}{{\partial {|r|^2}}}} \right)} \right) =  \\ - \frac{2}{3}\Lambda {e^{2a\left| r \right| + u}} -\frac{1}{3}{k^2}\left( {-{e^{u}}{\lambda _{tt}} - 2{\lambda _{xx}} +{\lambda _{yy}} + {e^{3 u}}{\lambda _{zz}} + {e^{2a|r|+ u}}{\lambda _{rr}}} \right), \\
\nonumber \\
\frac{1}{2}{e^u}\left( {\frac{{{\partial ^2}u}}{{\partial {t^2}}} - {e^{2a|r|}}\left( {8{a^2} + 4a \frac{{\partial u}}{{\partial {|r|}}}+2\delta \left( r \right)\frac{{\partial u}}{{\partial {|r|}}}  + 4a\delta \left( r \right) + \frac{{{\partial ^2}u}}{{\partial {|r|^2}}}} \right)} \right)=  \\  - \frac{2}{3}\Lambda {e^{2a\left| r \right| + u}} -\frac{1}{3}{k^2}\left( {{-e^{u}}{\lambda _{tt}} + {\lambda _{xx}} -2{\lambda _{yy}} +{e^{3u}}{\lambda _{zz}} +{e^{ 2a|r| + u}}{\lambda _{rr}}} \right), \\
\nonumber \\
{e^{ - 2u}}\left( { - \frac{{{\partial ^2}u}}{{\partial {t^2}}} + {e^{2a|r|}}\left( {\frac{{{\partial ^2}u}}{{\partial {|r|^2}}} + 4a\frac{{\partial u}}{{\partial {|r|}}} - 4{a^2} - 2a\delta \left( r \right)+2\delta \left( r \right)\frac{{\partial u}}{{\partial {|r|}}}  } \right)} \right) =  \\  - \frac{2}{3}\Lambda {e^{2a\left| r \right| - 2u}}  - \frac{1}{3}{k^2}\left( {{-e^{ -2u}}{\lambda _{tt}} + {e^{-3 u}}{\lambda _{xx}} + {e^{ -3 u}}{\lambda _{yy}} -2{\lambda _{zz}} +{e^{ 2a|r| - 2u}}{\lambda _{rr}}} \right), \\
\nonumber \\
 - \frac{3}{2}{\left( {\frac{{\partial u}}{{\partial |r|}}} \right)^2} - 4{a^2} - 8a\delta \left( r \right) =  - \frac{2}{3}\Lambda  - {k^2}{\left( {\frac{{\partial \phi }}{{\partial |r|}}} \right)^2} + \\ + \frac{1}{3}{k^2}\left( {{e^{ - 2a|r|}}{\lambda _{tt}} - {e^{ - 2a|r| - u}}{\lambda _{xx}} - {e^{ - 2a|r| - u}}{\lambda _{yy}} - {e^{ - 2a|r| + 2u}}{\lambda _{zz}} + 2{\lambda _{rr}}} \right), \\
\nonumber \\
- \frac{3}{2}~{\mathop{\rm sgn}} \left( r \right)~\frac{{\partial u}}{{\partial t}}\frac{{\partial u}}{{\partial |r|}} =  - {k^2}~{\mathop{\rm sgn}} \left( r \right)~\frac{{\partial \phi }}{{\partial t}}\frac{{\partial \phi }}{{\partial |r|}}.
\end{array}
\end{equation}

To solve this system of equations we split it as follows:
\begin{equation}\label{a-Lambda-relation}
\begin{array}{l}
\Lambda =6a^2,
\end{array}
\end{equation}
\begin{equation}\label{U-Phi-Relation}
\begin{array}{l}
 \frac{3}{2}{\left( {\frac{{\partial u}}{{\partial t}}} \right)^2}= {k^2}{\left( {\frac{{\partial \phi }}{{\partial t}}} \right)^2}, ~~~~~~
 \frac{3}{2}{\mathop{\rm sgn}} \left( r \right)\frac{{\partial u}}{{\partial t}}\frac{{\partial u}}{{\partial |r|}} =  {k^2}{\mathop{\rm sgn}} \left( r \right)\frac{{\partial \phi }}{{\partial t}}\frac{{\partial \phi }}{{\partial |r|}},
\end{array}
\end{equation}
\begin{equation}\label{U-equation}
\begin{array}{l}
 {\frac{{{\partial ^2}u}}{{\partial {t^2}}} - {e^{2a|r|}}\left( {4a \frac{{\partial u}}{{\partial {|r|}}}+ \frac{{{\partial ^2}u}}{{\partial {|r|^2}}}} \right)} = 0;
\end{array}
\end{equation}
\begin{equation}\label{SystemForBraneTensor}
\begin{array}{l}
2a\delta \left( r \right){e^{2a\left| r \right|}} = \frac{1}{3}{k^2}\left( {{2\lambda _{tt}} + {e^{- u}}{\lambda _{xx}} + {e^{- u}}{\lambda _{yy}} + {e^{2u}}{\lambda _{zz}} + {e^{ 2a|r|}}{\lambda _{rr}}} \right),\\
\delta\left( r \right) {e^{2a|r|+u}} \left( {2a+\frac{{\partial u}}{{\partial {|r|}}}} \right) = \frac{1}{3}{k^2}\left( {-{e^{u}}{\lambda _{tt}} - 2{\lambda _{xx}} +{\lambda _{yy}} + {e^{3 u}}{\lambda _{zz}} + {e^{2a|r|+ u}}{\lambda _{rr}}} \right), \\
\delta \left( r \right){e^{2a|r|+u}}\left( { 2a+\frac{{\partial u}}{{\partial {|r|}}}} \right)= \frac{1}{3}{k^2}\left( {{-e^{u}}{\lambda _{tt}} + {\lambda _{xx}} -2{\lambda _{yy}} +{e^{3u}}{\lambda _{zz}} +{e^{ 2a|r| + u}}{\lambda _{rr}}} \right), \\
2\delta \left( r \right){e^{2a|r| - 2u}}\left( {a-\frac{{\partial u}}{{\partial {|r|}}}  } \right) = \frac{1}{3}{k^2}\left( {{-e^{ -2u}}{\lambda _{tt}} + {e^{-3 u}}{\lambda _{xx}} + {e^{ -3 u}}{\lambda _{yy}} -2{\lambda _{zz}} +{e^{ 2a|r| - 2u}}{\lambda _{rr}}} \right), \\
 - 8a\delta \left( r \right) =   \frac{1}{3}{k^2}\left( {{e^{ - 2a|r|}}{\lambda _{tt}} - {e^{ - 2a|r| - u}}{\lambda _{xx}} - {e^{ - 2a|r| - u}}{\lambda _{yy}} - {e^{ - 2a|r| + 2u}}{\lambda _{zz}} + 2{\lambda _{rr}}} \right).
\end{array}
\end{equation}

 Equation \eqref{a-Lambda-relation} fixes relation between bulk cosmological constant $\Lambda$ and the parameter $a$ in the exponential warp factor of the metric \eqref{MetricAnsatz}, and through it the bulk cosmological constant will be fine-tuned to the brane tension \eqref{BraneTensor}.

 According to equation \eqref{U-Phi-Relation} we can set the relation between metric function $u\left(t,|r|\right)$ and phantom-like scalar field \begin{math}\phi\left(t,|r|\right) \end{math} as

 \begin{equation}
 u\left(t,|r|\right) = \sqrt{\frac{2k^2}{3}} \phi\left(t,|r|\right).
 \end{equation}

The system of five equations \eqref{SystemForBraneTensor} for the brane energy-momentum tensor components can be easily solved and the result is as follows:
\begin{equation}\label{BraneTensor}
\begin{array}{l}
\lambda_{tt}=-6ak^{-2}\delta \left(r\right),~~\lambda_{xx}=k^{-2}\left(6a-\frac{{\partial u}}{{\partial {|r|}}}\left(t,0\right)\right)e^{u\left(t,0\right)}\delta \left(r\right),~~\lambda_{yy}=k^{-2}\left(6a-\frac{{\partial u}}{{\partial {|r|}}}\left(t,0\right)\right)e^{u\left(t,0\right)}\delta \left(r\right),\\\\\lambda_{zz}=k^{-2}\left(6a+2\frac{{\partial u}}{{\partial {|r|}}}\left(t,0\right)\right)e^{-2u\left(t,0\right)}\delta \left(r\right),~~\lambda_{rr}=0,
\end{array}
\end{equation}
where $u\left(t,0\right)$ and $\frac{{\partial u}}{{\partial {|r|}}}\left(t,0\right)$ denote the values of $u(t,r)$ and its derivative with respect to $|r|$ at the origin $r=0$ respectively.

Now we return to equation \eqref{U-equation} and, using the separation of variables $u\left( {t,|r|} \right) = \tau \left( t \right)\lambda \left( |r| \right)$, decouple it as follows:

\begin{equation}\label{8}
\ddot \tau -C\tau=0,~~~~\lambda '' + 4a\lambda ' - C{e^{ - 2a|r| }}\lambda  = 0,
\end{equation}
where $C$ is some constant and overdots and primes denote derivatives with respect to $t$ and $|r|$, respectively.
There are three different cases for $C$, namely  $C=0$, $C=\omega^2$ and $C=-\omega^2$, where $\omega>0$ is some real constant.

For the case $C=0$ we have following equations:
\begin{equation}\label{9}
\ddot \tau  = 0,~~~~\lambda '' + 4a\lambda ' = 0,
\end{equation}
and the corresponding solution is:
\begin{equation}\label{10}
\tau \left( t \right) = {c_1} + {c_2}t,~~~~\lambda \left( |r| \right) = {c_3} + {c_4}{e^{ - 4a|r|}},
\end{equation}
where $c_1$, $c_2$, $c_3$ and $c_4$ are some real constants.

For the case $C=\omega^2>0$ equations get following form:
\begin{equation}\label{11}
\ddot \tau -\omega ^2\tau=0,~~~~~~\lambda '' + 4a\lambda ' - {\omega ^2}{e^{ - 2a|r|}}\lambda  = 0,
\end{equation}
and the corresponding solution is:
\begin{equation}\label{12}
\begin{array}{l}
\tau \left( t \right) = {c_1}{e^{\omega t}} + {c_2}{e^{ - \omega t}},\\
\lambda \left( |r| \right) = {c_3}e^{ - 2a|r|} I_2\left( \frac{{\omega }}{a}{e^{ - a|r|}} \right) + {c_4}e^{ - 2a|r|} K_2\left( \frac{{\omega }}{a}{e^{ - a|r|}} \right),
\end{array}
\end{equation}
where $c_1$, $c_2$, $c_3$ and $c_4$ are some real constants, $I_2\left(x\right)$ and $K_2\left(x\right)$ are second-order modified Bessel functions of first and second kind respectively.

And, finally, for the case $C=-\omega^2<0$ equations get form:
\begin{equation}\label{13}
\ddot \tau -\omega ^2\tau=0,~~~~~~\lambda '' + 4a\lambda ' + {\omega ^2}{e^{ - 2a|r|}}\lambda  = 0,
\end{equation}
and the corresponding solution is:
\begin{equation}\label{14}
\begin{array}{l}
\tau \left( t \right) = {c_1}\sin \left( {\omega t} \right) + {c_2}\cos \left( {\omega t} \right),\\
\lambda \left( |r| \right) = {c_3}e^{ - 2a|r|} J_2\left( \frac{{\omega }}{a}{e^{ - a|r|}} \right) + {c_4}e^{ - 2a|r|} Y_2\left( \frac{{\omega }}{a}{e^{ - a|r|}} \right),\end{array}
\end{equation}
where $c_1$, $c_2$, $c_3$ and $c_4$ are some real constants, $J_2\left(x\right)$ and $Y_2\left(x\right)$ are second-order Bessel functions of first and second kind respectively.

In the article \cite{gog-sing_2010} authors examined the solution (\ref{14}) and some its physical implications. In our future publications we will investigate trapping of various matter fields (see \cite{gog-mid1,gog-mid2,gog-mid3,gog-mid4,gog-mid5,gog-mid6}) on the brane at the origin of extra coordinate in the models defined by all this solutions.
%%%%%%%%%%%%%%%%%%%%%%%%%%%%%%%%%%%%%%%%%%%%%%%%%%%%%%%%%%%%%%%%

%%%%%%%%%%%%%%%%%%%%%%%%%%%%%%%%%%%%%%%%%%%%%%%%%%%%%%%%%%%%%%%%%%%%%%%%

%%%%%%%%%%%%%%%%%%%%%%%%%%%%%%%%%%%%%%%%%%%%%%%%%%%%%%%%%%%%%%%%%%%%%%%%

\medskip
\noindent {\bf Acknowledgments:}

The research of P. M. was supported by the grant of Shota Rustaveli National Science Foundation (former Georgian National Science Foundation") \#{\rm{GNSF}}/{\rm{ST}}0{\rm{9}}\_{\rm{798}}\_{\rm{4}} - {\rm{1}}00.

%%%%%%%%%%%%%%%%%%%%%%%%%%%%%%%%%%%%%%%%%%%%%%%%%%%%%%%%%%%%%%

\end{document}